\documentclass[10pt,preprint,a4paper]{aastex}
\usepackage{amsmath}                
\usepackage{amsfonts}               
\usepackage{amssymb}                
\usepackage{epsfig}                 
\usepackage{float}                  

\newcommand{\erg}{{~\rm erg}}
\newcommand{\yr}{{~\rm yr}}

\newcommand{\keV}{{~\rm keV}}

\def \astrobj#1{#1}

\begin{document}

\title{Core collapse supernova remnants with ears}

\author{Aldana Grichener\altaffilmark{1} and Noam Soker\altaffilmark{1}}

\altaffiltext{1}{Department of Physics, Technion -- Israel Institute of Technology, Haifa
32000, Israel; aldanag@campus.technion.ac.il; soker@physics.technion.ac.il}

\begin{abstract}
We study the morphologies of core collapse supernova remnants (CCSNRs) and find that about third of CCSNRs in our sample have two opposite `ears' protruding from their main shell. We  assume that the ears are formed by jets, and  argue that these properties are compatible with the expectation from the explosion jet feedback mechanism (JFM). Based on previous studies of ears in CCSNRs and the similarity of some ears to those found in planetary nebulae, we assume that the ears are inflated by jets that are launched during the explosion, or a short time after it. Under simple geometrical assumptions we find that the extra kinetic energy of the ears is in the range of 1 to 10 percents of the explosion energy. As not all of the kinetic energy of the jets ends in the ears, we estimate that the typical kinetic energy in the jets that inflated the ears,  under our assumptions,  is about 5 to 15 percents of the explosion energy. This study supports a serious consideration of jet-driven core-collapse supernova mechanisms. 
\end{abstract}


\section{INTRODUCTION}
\label{sec:intro}

In the explosion of core-collapse supernovae (CCSNe) a huge amount of gravitational energy is released when the inner part of the core collapses to form a Neutron Star (NS) or a Black Hole (BH).
There are two contesting proposed processes for channeling this gravitational energy to the  exploding star. The older one and much better studied is the delayed neutrino mechanism (\citealt{Wilson1985} and \citealt{BetheWilson1985}, and, e.g., \citealt{Jankaetal2016}, and  \citealt{Muller2016} for reviews).
The second one is the jet-feedback mechanism (JFM; \citealt{Soker2016}), where the jittering-jets mechanism (e.g., \citealt{PapishSoker2011, PapishSoker2014a, PapishSoker2014b, GilkisSoker2014, GilkisSoker2015, GilkisSoker2016}) operates in most regular CCSNe, and well collimated jets operate in super-energetic CCSNe \citep{Gilkisetal2016, Gilkis2016}. 
There is an alternative scenario where the explosion energy of massive stars originates in the nuclear burning of helium with oxygen \citep{Burbidgeetal1957}. However,  the rapid pre-collapse core rotation that is required for this scenario to work \citep{Kushnir2015} will lead to the launching of jets that carry much more energy than that released by the thermonuclear burning \citep{Gilkisetal2016}.

According to the JFM, when the jets have enough energy to unbind the core (considering that the efficiency is not 100 per cents), explosion takes place. The final activity episode of the jets takes place when the core has already been ejected. The final parts of the jets might expand freely to very large distances, interact with the already ejected core, and inflate the ears.
In cases of a slow pre-collapse core rotation the jets' axis wobble, and the explosion is termed the jittering jets explosion mechanism (e.g., \citealt{PapishSoker2014a, PapishSoker2014b}). There are several to few tens of jet-launching episodes in different directions. In each launching episode the jets carry an amount of energy of few to about few tens percents of the explosion energy. The jets of the last episode, that are launched after the ejection of the core, might inflate ears. 
\cite{Castellettietal2006} suggest that the SNR Puppis A with a morphology that contains two opposite `ears', was shaped by jets.
Such jets might be the last two opposite jets launched in the jittering jets mechanism, or the signature of well collimated long-lived double jets.
Two opposite ears are observed also in Type Ia SNR (SNR Ia). \cite{TsebrenkoSoker4472015} list such SNR Ia, and study the formation of such opposite ears by jets.

It might be possible that the ears are formed shortly after the explosion by jets launched from a rapidly rotating NS remnant. However, the formation of a rapidly rotating NS is most likely to be accompanied by the launching of jets \citep{Soker2016a, Soker2017}, hence bringing us back to explosion by jets.

The JFM is not yet well tested, and it is still in a speculative stage of development. In particular, there are the questions of the source of angular momentum of the accreted mass to form an accretion disk (or belt), the launching process of the jets, nucleosynthesis in the explosion, and the energy of the explosion. The very high spatial resolution that is required in three dimensions magneto-hydrodynamical simulations slows down the study of these processes. 

There are three lines of arguments to support the operation of the JFM in the explosion of massive stars \citep{Soker2016}.  (1) Observations show that most CCSNe explode with a typical energy (mainly the kinetic energy), that is about equal to and up to several times the binding energy of the ejected mass, $E_{\rm explosion} \simeq {\rm few} \times E_{\rm bind} \approx 10^{51} \erg$. This hints at the operation of a negative feedback mechanism.

(2) Jets and asymmetry. (2.1) The presence of jets in long gamma ray bursts (GRBs; e.g., \citealt{Woosley1993, ShavivDar1995, SariPiran1997}). Some long GRBs are observed to be associated with  Type Ic supernovae (e.g., \citealt{Canoetal2016}), showing that jets can be produced at the collapse.
(2.2) Jets in CCSNRs. There are claims for a real jet in the SN remnant (SNR) {Cassiopeia~A}
\citep{FesenMilisavljevic2016}, and observations for a significant role played by jets in at least some CCSNe  (e.g. \citealt{Lopezetal2011, Lopezetal2013, Lopezetal2014, Milisavljevic2013, Gonzalezetal2014}). 
(2.3) Polarization measurements. Other supports come from polarization measurements which show departure from axial symmetry (e.g., \citealt{Wangetal2001, Maundetal2007}). \cite{Maundetal2007} attribute the polarization in the Type Ib/c SN 2005bf to a  Nickel-56 rich jet whose axis is tilted with respect to the axis of the photosphere, that has penetrated the C–O core, but not the He mantle. The jittering-jets mechanism might account for such jets. Polarimetric observations of SN~2015bn that indicate an elongated morphology further support the presence of jets in CCSNe \citep{Inserraetal2016}.

(3) The 32 years old delayed neutrino mechanism encounters three problems  \citep{Papishetal2015}.
($i$) In the delayed-neutrino mechanism the explosion starts with the revival of the stalled shock of the inflowing core gas. This is not always achieved in numerical simulations, even in the most sophisticated ones. ($ii$) Even if the stalled shock is revived, in most simulations the desired energy of $\approx 10^{51} \erg$ is not achieved. ($iii$) When in simulations based on neutrino-driven explosion the explosion energy is scaled to observed CCSNe, such as \astrobj{SN~1987A}, the maximum energy that the delayed-neutrino mechanism can supply is about $2 \times 10^{51} \erg$ (e.g., \citealt{Sukhboldetal2016, SukhboldWoosley2016}). Even when convection-enhanced neutrino-driven explosion is considered (convective-engine) the explosion energy cannot get higher than this limit \citep{Fryeretal2012}. Therefore, the delayed neutrino mechanism cannot account for super energetic CCSNe (SESNe).

Jet-driven explosions of CCSNe have been studied over the years (e.g. \citealt{LeBlanc1970, Meier1976, Bisnovatyi1976, Khokhlov1999, MacFadyen2001, Hoflich2001, Woosley2005, Burrows2007, Couch2009, Couch2011, TakiwakiKotake2011, Lazzati2012, Mostaetal2014, Mostaetal2015, BrombergTchekhovskoy2016}).
The constant direction of the jets in these studies requires that there will be a well defined angular momentum axis, implying that the pre-collapsing core has a rapid rotation. These are rare cases, and they do not operate via an efficient JFM.
It seems that jets are more common than what the initial conditions in these studies require, and more in line with the JFM for exploding CCSNe in our sample.

\section{THE RELATION BETWEEN JETS AND EARS}
\label{sec:jets}

A protrusion from the main SNR shell that possesses one or more of the following properties will be referred  to as an `ear'. (1) It has a different characteristics than the main SNR shell. Examples include a substantially different ratio of radio to X-ray surface brightness, or that the ear is disconnected from the main SNR shell by a faint region. (2) In a case that the SNR shell has a general elliptical appearance, the protrusion is along the long axis of the SNR. (3) There are two opposite protrusions with respect to the center of the SNR.

 Let us elaborate on these criteria. As to criterion (1). In many cases one of the different characteristics of the ears is that they are fainter than the rest of the nebula. Therefore, the faint ears will reveal themselves mainly in the most sensitive wavelengths. The best wavelength bands might change from one SNR to another, much as the main shells of different SNRs are revealed at different wavelengths, depending on the evolutionary state of the SNR.
 In particular, wavelengths that reveal the inner regions of the SNR will not reveal the ears that are on the outskirts of the SNR. Take the CCSNR G309.2-00.60 as one example. Its radio image reveal the entire shell and the ears \citep{Gaensleretal1998}, while its X-ray image barely reveals the main shell, yet not at all the ears \citep{Rakowskietal2001}.
Another example is the Vela CCSNR, where the X-ray image shows eminent ears, while the radio image shows the filamentary nature of the remnant, but does not reveal the ears \citep{Bocketal1998}.

Criteria (2) and (3) are the expected morphologies of ears formed by jets, as evident from many planetary nebulae (e.g., \citealt{Balick1987}). These criteria have another advantage. The appearance of ears along the major axis and having two ears on opposite sides are properties that are very unlikely to be caused by interaction with the ISM. Later we will use these properties to attribute the presence of ears even in very old CCSNRs to the energetic jets that shaped the ears, rather to an interaction with the ISM. 

There are three types of interactions that can in principle lead to the formation of two opposite ears by jets in the SNR. Note that in many cases we might observe only one of the two ears. (1) Pre-explosion ears in the circumstellar matter (CSM) in the near vicinity of the progenitor. The spherical explosion gas fills the CSM shell, and the imprint of ears survives to the SNR phase. This can be the case if the progenitor is in a binary system, and a mass transfer leads the mass-accreting companion to launch jets. These jets form ears in the CSM before explosion. (2) Jets that are launched during the explosion.
(3) Post-explosion jets that are launched by the remnant, either a NS or a BH. These jets can be launched shortly after the explosion or a long time after, e.g., in a PWN. For example, in a recent paper \cite{Olmietal2016} conducted a 3D simulation of a PWN and demonstrated the formation of jets. In all three scenarios the jets are not active during the SNR phase, unless a close binary companion transfers mass to the compact remnant (see below).

Another possibility of an axisymmetrical pre-explosion CSM was simulated by \cite{Blondinetal1996}, and involves no jets. They set a spherical explosion inside a CSM that has a density decreasing toward the poles. They manged to obtain ears. For that, it is possible that some ears are form by such an interaction, rather than by jets. We here assume that the ears are formed by jets.  

In SN Ia the exploding white dwarf (WD) does not survive. Hence, only the first two possibilities might account for ears in SNR Ia. \cite{TsebrenkoSoker2013} conducted hydrodynamical simulations and demonstrated that both pre-explosion and during-explosion jets can in principle form the ears that are observed in SNR Ia. It is not clear yet how an exploding WD can launch jets, i.e., it might require a rapidly rotating WD. For that, \cite{TsebrenkoSoker4472015} and \cite{TsebrenkoSoker4502015} prefer the pre-explosion ear scenario to account for ears in SNR Ia.

In CCSNe all three types of processes for the formation of ears can in principle operate.  The bipolar structure of the three rings around SN 1987A demonstrates that the CSM of a CCSN can have an axisymmetrical structure, most likely formed by a strong binary interaction that involves a merger process (e.g., \citealt{MorrisPodsiadlowski2009}).
In the specific case of SN~1987A the binary system merged long before the explosion (e.g., \citealt{ChevalierSoker1989, Podsiadlowskietal1990}).
In the case of a merger, the pre-explosion star is likely to posses a relatively rapid rotation. This increases the likelihood of the formation of an accretion disk around the newly born NS, and hence of the launching of jets (e.g., \citealt{Gilkis2016}).

\begin{figure}[H]
\centering
\includegraphics[width=0.8\textwidth]{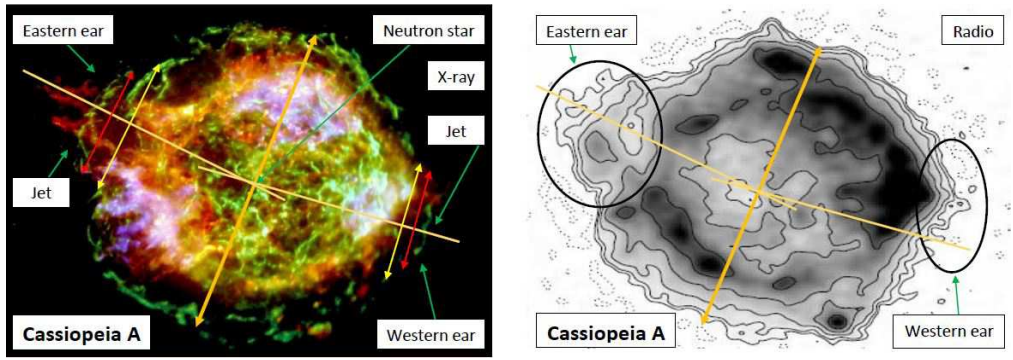}
\caption{Images of the SNR Cassiopeia A. Left: An X-ray image taken from the chandra gallery (based on \citealt{Hwangetal2004}). Red, blue and green represent Si He$\alpha$ (1.78-2.0 keV), Fe K (6.52-6.95 keV), and  4.2-6.4 keV continuum, respectively. 
Right: A radio image at 69 MHz obtained from a single 0.2 MHz sub-band, as published by \cite{Oonketal2016}. We added arrows to indicate different quantities that we will use in our analysis in section \ref{sec:energy}.
}
\label{fig:CassiopeiaA}
\end{figure}

We show the X-ray and radio images of SNR Cassiopeia~A in Fig. \ref{fig:CassiopeiaA}.
Only the eastern jet and ear are clearly seen. There is a Si-rich counter jet more or less where we mark the western ear \citep{Hwangetal2004}, that supports the notion that there is a counter ear. They comment that the bipolar structure is due to jets during the explosion, rather than to pre-explosion cavities in the CSM.

The X-ray image presents a bright outer region (in green) which marks the location of a shock wave generated by the SNR expansion. The eastern ear is clearly seen by this thin green region, and the radio image. 
  
\begin{table}[H]
\begin{center}
 \begin{tabular}{||l c c c l||}
 \hline
 SNR & Figure & Emission band & Wavelengths/Energy Band & Source \\ [0.5ex]
 \hline
 Cassiopeia A & 1 (right) & radio & 4.34 [m] & \cite{Oonketal2016} \\
 Cassiopeia A & 1 (left) & X-Ray & 1.78-2.0, 4.2-6.4, 6.52-6.95 [keV] & \cite{Hwangetal2004}  \\
 3C58 & 3 & composite & 0.5-10 [keV] & \cite{Slaneetal2004} \\
 Puppis A & 4 (right) & X-Ray & 0.3-8.0 [keV] & \cite{Dubneretal2013} \\
 Puppis A & 4 (left) & radio & 0.214 [m] & \cite{ReynosoWalsh2015}  \\
 S147 & 5 & H$\alpha$ & 656 [nm] & \cite{Drewetal2005}  \\
 Vela & 6 & composite & 0.1-2.4 [keV] & \cite{Aschenbachetal1995}  \\
 G309.2-00.6 & 7 & radio & 0.231 [m] & \cite{Gaensleretal1998} \\
 W44 & 8 & composite & 0.7-2.6 [keV], 4.5 [$\mu$m] & \citealt{Sheltonetal2004}, Spitzer\\
 Crab Nebula & 9 & composite & 0.4-2.1 [keV],visible,4.5 [$\mu$m]  & \cite{Sewardetal2006},\cite{Hester2008} \\ [1ex]
 \hline
\end{tabular}
\centering
\caption{ Some information on the presented images of CCSNe.
The ear-like features appear in images in the radio, IR, visible, and X-ray, and are not unique to a specific emission mechanism.  
}
\label{table:summary}
\end{center}
\end{table}
    
The eastern jet-ear structure of the SNR Cassiopeia~A clearly demonstrates the relation between a jet that was launched during the explosion \citep{Hwangetal2004, Lamingetal2006} and the presence of an Ear. We can also note the ears in the detailed high-resolution VLA radio images at 1.4 and 4.8 GHz studied by \cite{AndersonRudnick1995}.
In Cassiopeia~A the jet penetrated the SNR shell and its head precedes the ear. In some other cases the jet might not penetrate the SNR shell, and its material is enclosed within the ear and SNR shell. If we are to learn from SNR Cassiopeia~A, then the presence of ears supports the significant role that jets play in the explosion mechanism of CCSNe.

\cite{DeLaneyetal2010} study the 3D structure of Cassiopeia~A, and argue that the structure was formed by the explosion itself rather than interaction with ambient gas (see also \citealt{Lamingetal2006}). They do not find a large scale bipolar-structure, but rather multiple outflows (which they termed pistons), including a bipolar structure that has the jet and a counter jet in it. We identify the two ears at the two sides of their bipolar structure.  

 Despite the discussion we have made so far, we do note that there are alternative explanations for the formation of the ears in some SNRs, such as the attribution of the ears in W50 and in the Crab Nebula to magnetic fields (e.g., \citealt{BegelmanLi1992}), or the interpretation of the asymmetries of the SNRs Puppis A and S147 as by-products of the interaction of these SNRs with their environment \citep{ReynosoWalsh2015, Ngetal2007}. Below we will further mention alternative scenarios as we discuss specific CCSNRs. 

\begin{figure}[H]
\centering
\includegraphics[width=0.4\textwidth]{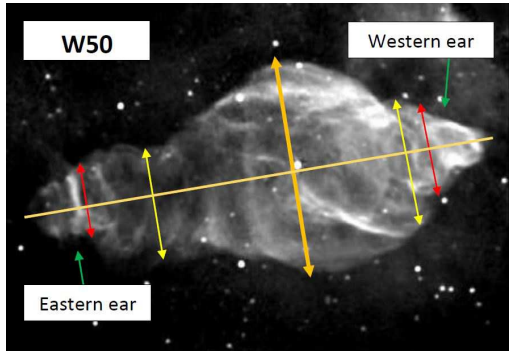}
\caption{The SNR W50 in radio continuum at 1415 MHz as observed with the VLA (from \citealt{Dubneretal1998}). We added arrows to indicate different characteristic quantities of the SNR that we define in section \ref{sec:ears}.
}
\label{figure:W50}
\end{figure}

The X-ray binary system SS433 and its associated SNR W50 (SNR G039.7-02.0) is a whole different story. Although it does not belong to the type of objects we study, we present it here to further emphasize the relation between jets that are launched by a NS or a BH, and the presence of ears.
In this system a binary companion transfers mass to the NS (or BH) remnant of the CCSN, and the accretion disc around the compact object launches relativistic jets. As can be seen in Fig. \ref{figure:W50}, the two ears occupy much of the volume of this SNR.
Most of the power from the jet seems to go into the mechanical energy, producing the ears \citep{SafiHarbPetre1999}.

In the systems we study there is no mass transfer from a companion. Nonetheless, the general structure of the ears in W50 is similar to the ears in the SNR G309.2-00.6 that we study later on \citep{Gaensleretal1998}. This suggests that the ears in SNR G309.2-00.6 were formed by jets that were launched during the explosion, or shortly after the explosion, as there are no active jets there anymore.

The main point of this section is the strong relation between the operation of jets, at present or in the past, and the presence of ears in SNRs. The morphologies of the ears we study, namely, that they are distinguished from the main SNR shell in a clear way, suggest to us that most of the ears where blown from the SNR shell, and are not just an elongation of the SNR.

As we cannot study the operation of jets in the past, we turn to study the ears that are observed in SNRs, and the implication of their properties on the jets that formed them and on the explosion mechanism of CCSNe.
 
\section{SUPERNOVA REMNANTS WITH EARS}
\label{sec:ears}

In this section, we present images of CCSNRs with ears. We looked at images of numerous CCSNRs in various wavelengths, presented in the Chandra (X-ray) gallery and in the Hubble (Optical) website. We examined profoundly 41 well studied CCSNRs in the Milky Way and in the Magellanic clouds, by conducting a more comprehensive search in the professional literature. Out of these 41 CCSNRs, for 28 (not including W50) the images are clear enough to tell whether ears exist or not. Information about the images presented in sections \ref{sec:jets} and \ref{sec:ears} is given in Table \ref{table:summary}. 

 We took the identification of the SNRs as CCSNRs from the literature. The studies we have used for the images of the SNRs identify them as CCSNRs. Out of the 41 SNRs we have examined, 19 have NS. When a NS is not observed, the identification is based on other expected properties of CCSNe, such as the relative abundance of iron. However, clear differences between CCSNRs and Type Ia SNRs exist only when the remnant is young (e.g., \citealt{Raymond1984, Lopezetal2011}). The identification is not always simple, as was demonstrated by the only recent classification of the SNR G344.7−0.1 as a CCSNR \citep{Lopezetal2011}.

Out of the 8 CCSNR that contain ears, 7 have central NS. Only the SNR G309.2-00.6 has no observed NS, and it was classified as a CCSNR by \cite{Gaensleretal1998} based on its morphology. This classification was solidified by \cite{Rakowskietal2001} based on the low iron abundance and the presence of significant neon and magnesium. 
We conclude that the identifications of the 8 CCSNRs we discuss in this study are robust (9 including W50). For the remaining 19 CCSNRs of the group of these 28 CCSNRs, for which the images were clear enough to find ears, 11 have NS. Other indications were used for the rest. The SNR Kes~79 (G33.6+0.1) is known to have a NS, hence it is a clear case of a CCSNR. In addition, \cite{Satoetal2016} argue that the abundance pattern and mass of the non-equilibrium ionization plasma indicate that the SNR originates from the CCSN with a $∼30-40 M_\odot$ mass progenitor. Namely, the abundances can also be used to indicate that a SNR comes from a CCSN.  
For the SNR~B0049–73.6 that has no central NS, the high abundance of O, Ne, Mg, and Si is a clear indication for CCSNR (e.g., \citealt{Schencketal2014}). 
In any case, if some of these 19 are not CCSNRs, but rather Type Ia SNRs, then the conclusions of our study are stronger even.

 We found that eight CCSNRs (not including the CCSNR W50) have distinct ears, while another four CCSNRs might have ears. Seven of CCSNR who posses ears have 2 prominent ears, and only one of them (the Crab Nebula) has one ear. If the symmetry axis, i.e., the line joining the two counter ears, is close to our line of sight, the ears will be projected on the main SNR shell and we might not see them. Therefore, the true fraction of CCSNRs with ears might be larger than the fraction of $8/28=29 \%$ to $12/28=43 \%$ we find here. At this stage we can say that at least about third of CCSNRs possess one or two ears.

Under the assumption that ears were shaped by jets, we set a goal to calculate the extra energy that was required to form each ear. For that, we define some geometrical properties of the ears as follows.
Although the ears are not spherical, we define a diameter for each ear, and mark it with
a double-headed red arrow. If a symmetry axis exists, the double-headed red arrow is more or less perpendicular to this axis.
We define the base of the ear as the region on the SNR main shell from which the ear protrude.
We cannot tell from the images what the shape of the base is. We assume it is circular.
With a yellow double-headed arrow we mark the diameter of the base of the ear on the SNR main shell. There are large uncertainties in the exact values of the diameters of the ears and bases. Nonetheless, we think they are adequately defined for our purpose of estimating the extra energy that is needed to inflate the ears. This calculation is postponed to section \ref{sec:energy}.
We also define a diameter for the SNR shell, or a typical size if the SNR substantially departs from a spherical geometry, and mark it with a double-headed thick orange arrow. In some cases we also mark the symmetry axis of the SNR.

In most images we were consistent with the colors of the arrows mentioned above. Nevertheless, in some cases, in order to stress the arrows over the background, we used different colors. In the image of Vela SNR we marked the radius of the ear with a double-headed purple arrow, the radius of the base with a double-headed brown arrow and the radius of the SNR itself with a double-headed thick blue arrow. In the image of SNR G309.2-00.6, we marked the base radius of the overlapping area with a double-headed bourdeaux arrow, and in the image of Puppis A we marked the radius of the eastern ear and it's base radius by using purple and black double-headed thick arrows, respectively. 

\begin{figure}[H]
\centering
\includegraphics[width=0.5\textwidth]{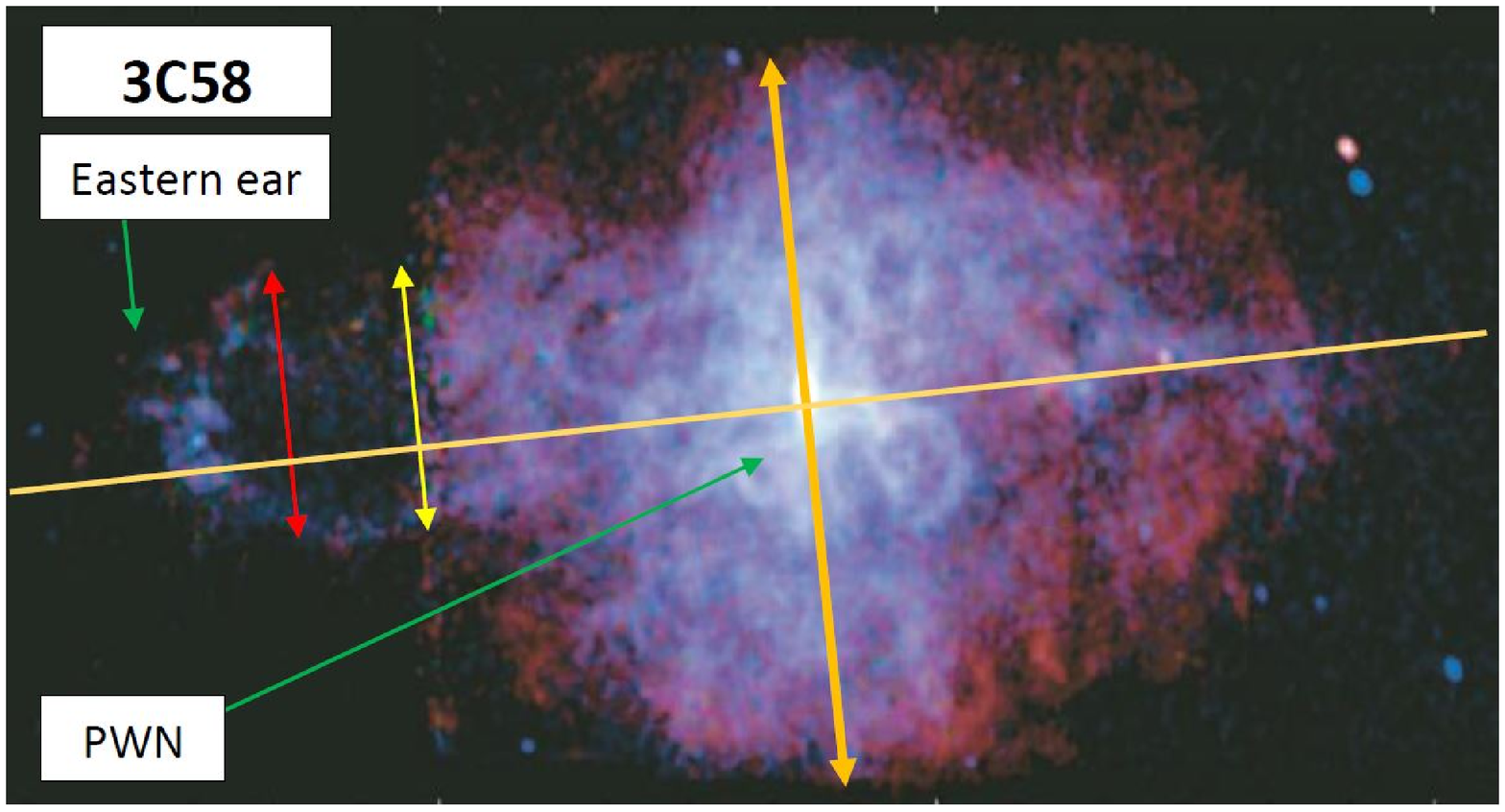}
\includegraphics[width=0.4\textwidth]{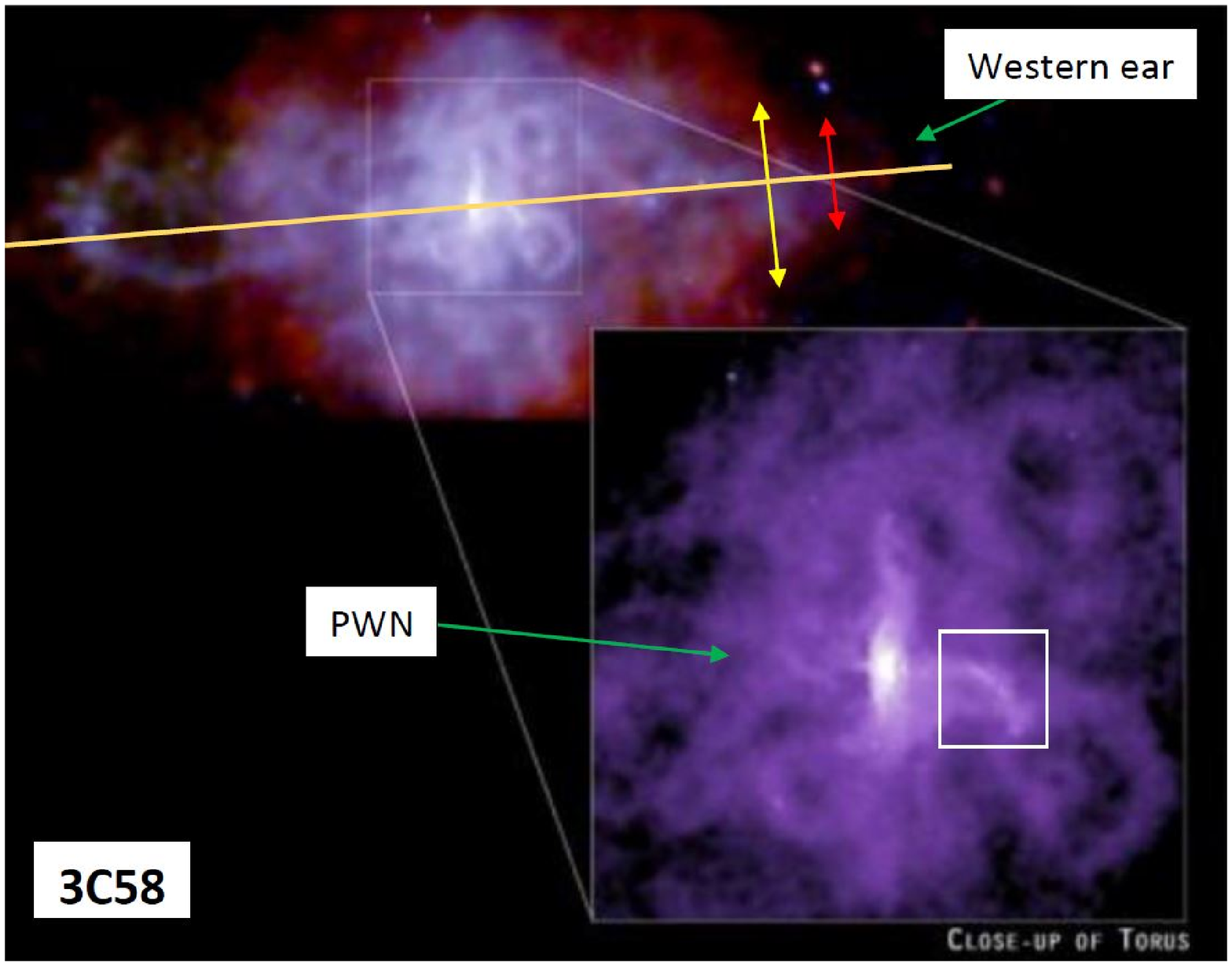}
\caption{The SNR 3C58. ACIS/Chandra images of 3C58 in the energy bands $0.5 - 1.0 \keV$ (red), $1.0-1.5 \keV$ (green), and $1.5-10 \keV$ (blue). Both panels are based on \cite{Slaneetal2004}; right image was taken directly from the Chandra gallery. The images do not have the same scale. 
}
\label{fig:3C58}
\end{figure}

{\bf SNR 3C58.} There is a rich literature on this SNR (e.g., \citealt{Slaneetal2004,LopezCotoetal2016}), yet we limit our study to the morphology of the ears. Its X-ray image is presented in Fig.~\ref{fig:3C58}, together with its PWN. This CCSNR has two opposite ears on the long axis of the SNR. The eastern ear is significantly larger and clearer than the western ear. We present two panels to better mark morphological parameters for each ear (the diameter of the ear and the diameter of the ear's base). Radio images of 3C58 also reveal the ears, although they are less pronounced in the radio map \citep{Bietenholz2006}.
Like seven out of the 8 CCSNe analyzed here (as presented in Table \ref{table:CC}), this CCSN has a central neutron star, the pulsar J0205+6449. It launches two opposite jets along the symmetry axis (e.g., \citealt{Shibanovetal2008}). 
\begin{table}[H]
\begin{center}
 \begin{tabular}{||l c l l||}
 \hline
 SNR & Central object & Name & Source \\ [0.5ex]
 \hline
 Cassiopeia A & NS & CAS A NS & \cite{DeLaneySatterfield2013} \\
 3C58 & Pulsar, PWN & PSR J0205+6449 & \cite{ShearerNeustroev2008}\\
 Puppis A & NS & RX J0822-4300 & \cite{HuiBecker2007} \\
 S147 & Pulsar, PWN & PSR J0538+2817 & \cite{Andersonetal1996}, \cite{RomaniNg2003} \\ 
 Vela & Pulsar, PWN &  PSR B0833-45 & \cite{Murdin2000}, \cite{Abramowskietal2012} \\
 G309.2-00.6 & Not detected & $-$ & \cite{Gaensleretal1998} \\
 W44 & Pulsar, PWN & PSR B1853+01 & \cite{Harrusetal1996}, \cite{Sheltonetal2004} \\
 Crab Nebula & Pulsar, PWN & PSR B0531+21 & \cite{Germanetal2012}, \cite{Kargaltsev2012} \\ [1ex]
 \hline
\end{tabular}
\centering
\caption{
 The names of the central NS in seven of the systems analyzed here. Only the CCSNR G309.2-00.6 has no detected NS, yet the data on this SNR make a strong case that it's appearance is significantly affected by collimated outflows from a central source \citep{Gaensleretal1998}. } 
 
\label{table:CC}
\end{center}
\end{table}

\begin{figure}[H]
\centering
\includegraphics[width=0.9\textwidth]{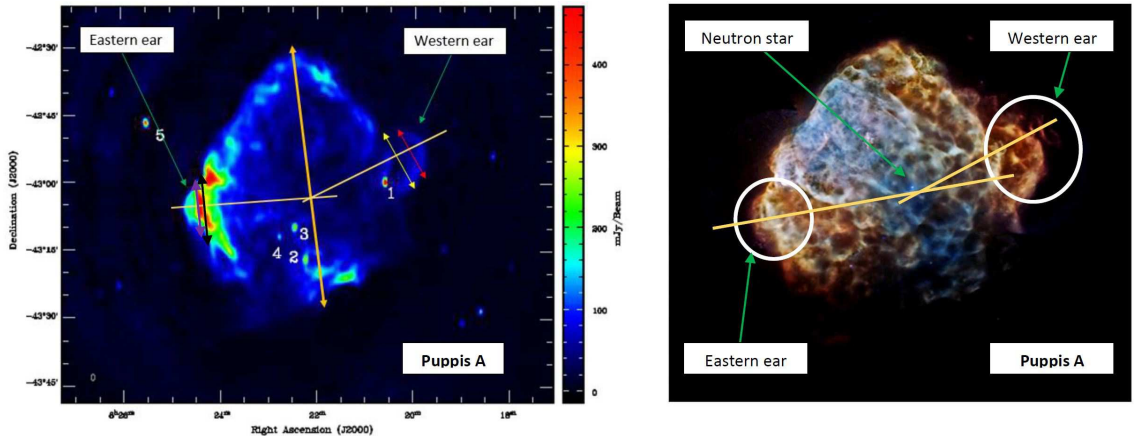}
\caption{The SNR Puppis~A. Left: The image displays the radio continuum emission produced at 1.4 GHz. It was published by \cite{ReynosoWalsh2015}, and reproduced by \cite{Reynosoetal2017}. Right: A full view of Puppis A in X-ray, where red, green, and blue correspond to the 0.3-0.7, 0.7-1.0, and 1.0-8.0 keV bands, respectively (from \citealt{Dubneretal2013}).
}
\label{figure:PuppisA}
\end{figure}

{\bf SNR Puppis~A. } In Fig.~\ref{figure:PuppisA} we present the well studied (e.g., \citealt{Dubneretal2013}) oxygen rich SNR Puppis A. The two small ears are clearly resolved in these images, and despite that the SNR has no symmetry axis, the two opposite ears are prominent. \cite{Castellettietal2006} and  \cite{DubnerGiacani2015} already mentioned that the ears could have been formed by jets.
\cite{ReynosoWalsh2015} suggested that the asymmetry of the two ears results from asymmetrical interaction with the ISM, and the easternmost filament may be the result of interaction of the SNR with an external cloud.
Because of the lack of symmetry, we draw a separate `symmetry-axis' for each ear.

\begin{figure}[H]
\centering
\includegraphics[width=0.55\textwidth]{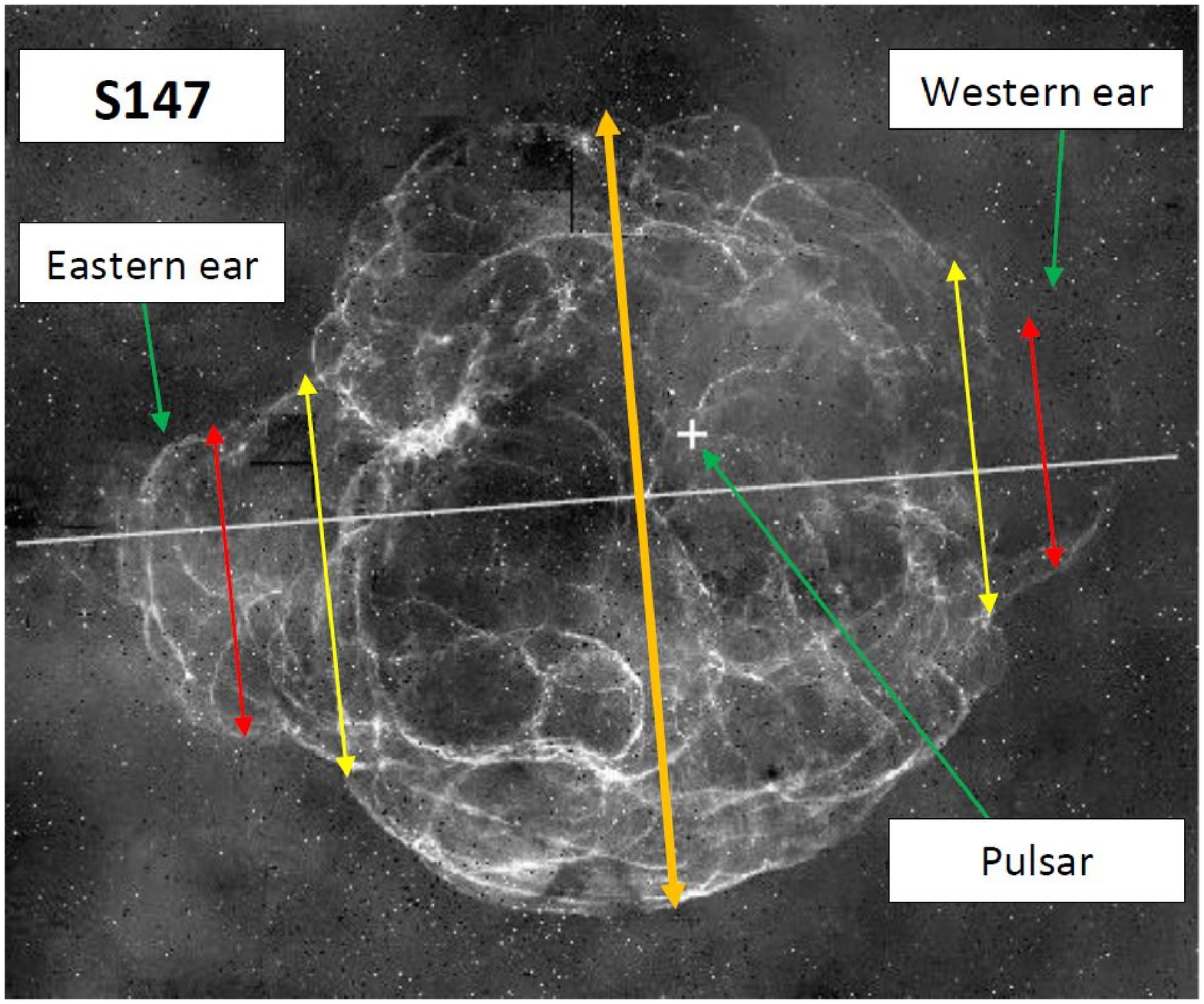}
\caption{An H$\alpha$ image of the SNR Semeis~147 taken from \cite{Gvaramadze2006} who reproduced an image from \cite{Drewetal2005}.
 The symmetry axis is from the original image.
}
 \label{figure:Semeis147}
\end{figure}

{\bf SNR Semeis~147.} In Fig.~\ref{figure:Semeis147} we present the H$\alpha$ image of Semeis~147. The shell has a general circular morphology, but with obvious blow-outs in the eastern and western directions, i.e., the ears. Those elongations define a bilateral axis passing through the center (e.g., \citealt{Ngetal2007, Gvaramadze2006}).The ears are seen also in radio images, but they are fainter relative to the main shell \citep{Xiaoetal2008}. The SNR contains a central pulsar, PSR J0538+2817.
\cite{Ngetal2007} attribute the asymmetrical expansion of this SNR to inhomogeneities in the surrounding interstellar medium. \cite{Gvaramadze2006} claims that a possible explanation to the extended morphology is that Semeis~147 is a remnant of a SN which exploded within a low density bubble surrounded by a shell. The bubble was formed by the stellar wind of the SN progenitor during its WR phase of evolution.  \cite{Dinceletal2015} raise the possibility that the elongation of Semeis~147 is due to the jets or the torus of the PWN. \cite{Dinceletal2015} further suggest that the progenitor was an interacting binary system. As we discussed in section \ref{sec:jets}, such a binary interaction might form pre-explosion ears. In this study we assume that the ears were formed by jets launched during the explosion or shortly after the explosion.

\begin{figure}[H]
\centering
\includegraphics[width=0.55\textwidth]{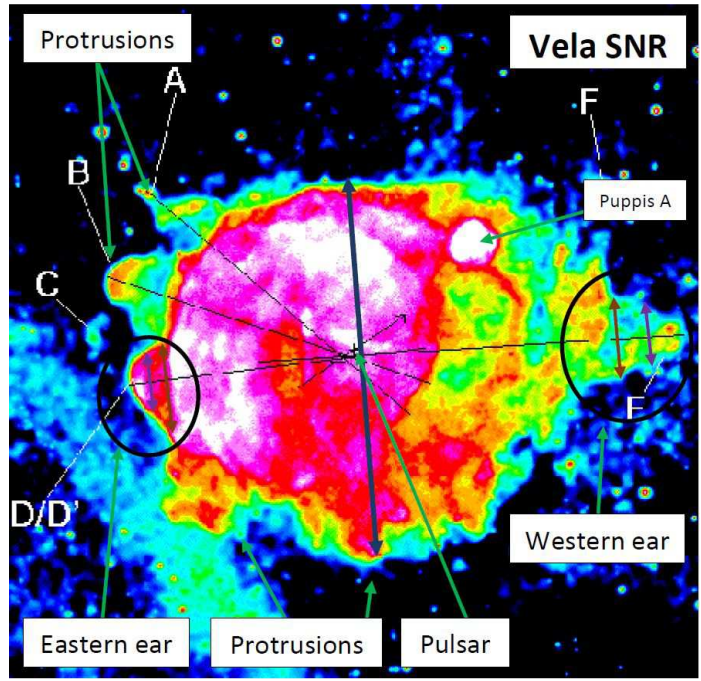}
\caption{The Vela SNR. This is a ROSAT all-sky survey image ($0.1 - 2.4 \keV$) from  \cite{Aschenbachetal1995}. The black lines, including the symmetry axis (the longest one), and the letters marks on the protrusions are from there as well.
We added the black ellipses to mark the location of the ears.
The two double-headed arrows in purple mark the diameter of the ears, while the blue-thick double-headed arrow marks the diameter of the SNR shell. The two double-headed brown arrows represent the diameter of the base of the ears. }
 \label{figure:Vela}
\end{figure}

{\bf The Vela SNR.} In Fig.~\ref{figure:Vela} we present the Vela SNR. The Vela SNR possesses many protrusions from its main shell. \cite{Aschenbachetal1995} labeled some of these protrusions, as can be seen in the figure. We added the black ellipses to mark the ears and the morphological quantities that are relevant to our study.
The bright clump is a separate SNR, Puppis A.
\cite{Micelietal2013} refer to the protrusions as detached clumps, and simulate the formation of the eastern ear with a dense clump that was formed during the explosion (i.e., not with a pulsar wind). The kinetic energy of the dense clump in their simulations is about $5 \%$ of the kinetic energy of the SNR shell. We accept their general view on the formation of the eastern ear, but attribute the formation of ears to jets, rather than to clumps. The outcome of a jet and a clump long after the jet has ceased is similar.
\cite{Gvaramadze1999}, on the other hand, attribute the structure of the Vela SNR to the interaction of the SN ejecta with a bubble blown prior to the explosion, accompanied by Rayleigh-Taylor instabilities.
The Vela SNR contains a central pulsar (PSR B0833-45; marked with a small black cross) and a long collimated jet-like structure (e.g., \citealt{Grondinetal2013}).

\begin{figure}[H]
\centering
\includegraphics[height=0.5\textwidth]{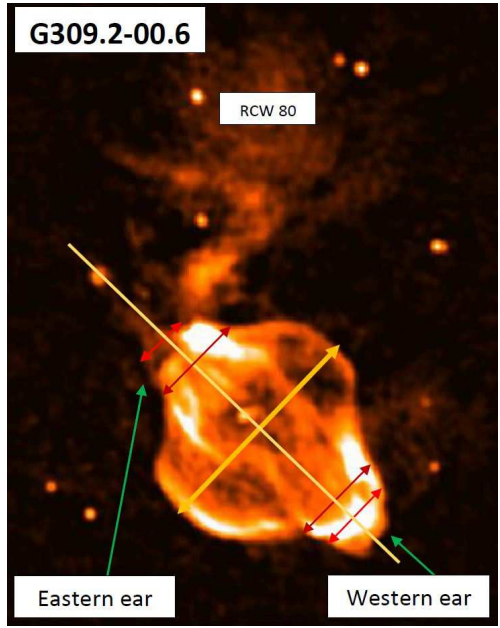}
\caption{The SNR G309.2-00.6. In the background lies the emission nebula RCW 80. The radio image is taken from the site of the School of Physics, The university of Sydney, where it was posted as production from \cite{Gaensleretal1998} 
Although no central NS has been detected, its morphology and location in the Galaxy strongly suggest that it is a CCSNR \citep{Gaensleretal1998}.
}
 \label{figure:G3092006}
\end{figure}

{\bf SNR G309.2-00.6.}
We present the radio image of this SNR in Fig.~\ref{figure:G3092006}. The geometrical quantities relevant to our study are marked on the figure.
This CCSNR has two clear ears as we mark on Fig.~\ref{figure:G3092006}. \cite{Gaensleretal1998} note the similarity between the ears in this SNR and those in SNR W50, and discuss the formation of the ears by jets launched by the remnant. Despite the resemblance between the morphologies of SNR G309.2-00.6 and SNR W50, there is no evidence of a binary companion to the remnant of SNR G309.2-00.6 and no evidence for jets \citep{SafiHarbetal2007}.

\begin{figure}[H]
\centering
\includegraphics[width=0.55\textwidth]{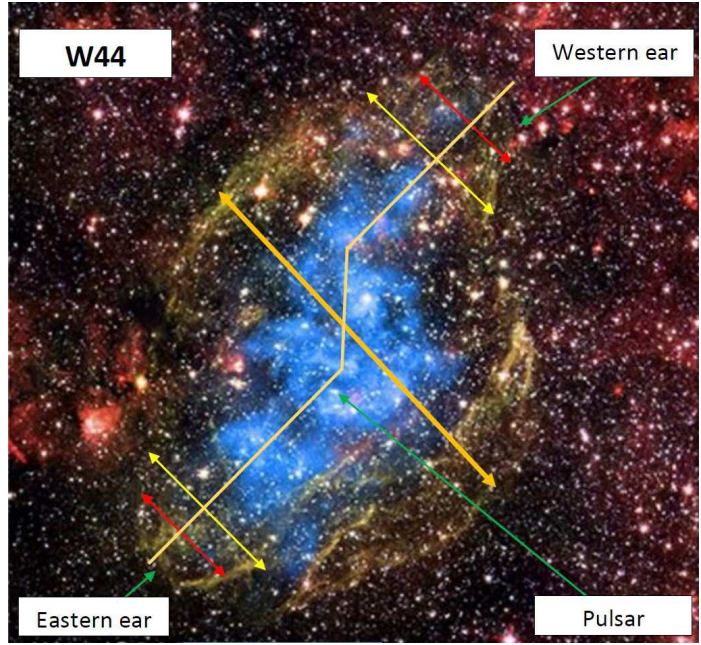}
\caption{The SNR W44. Composite image taken from the Chandra gallery. The cyan represents X-ray (based on \citealt{Sheltonetal2004}), while the red, blue and green represent infra-red (based on NASA/JPL-Caltech). We added three beige thick lines to schematically define the S-shape of this SNR. 
}
 \label{figure:W44}
\end{figure}

{\bf SNR W44. }
The image of SNR W44 is presented in Fig.~\ref{figure:W44}, together with our drawing of the relevant geometrical quantities of the ears. The ears are revealed in the infra-red, but neither in the X-ray (the inner bright region in cyan) nor in the radio \citep{Castelletti1etal2007, Egronetal2017}. W44 contains a central pulsar, PSR B1853+01, which has a PWN that emits radio and X-ray (e.g., \citealt{Sheltonetal2004}).
This SNR has the appearance of a non-circular shell elongated in the south-east and north-west directions (e.g., \citealt{Tanaka2009, Castelletti1etal2007}). Those elongations are the ears that we study.

\begin{figure}[H]
\centering
\includegraphics[width=0.55\textwidth]{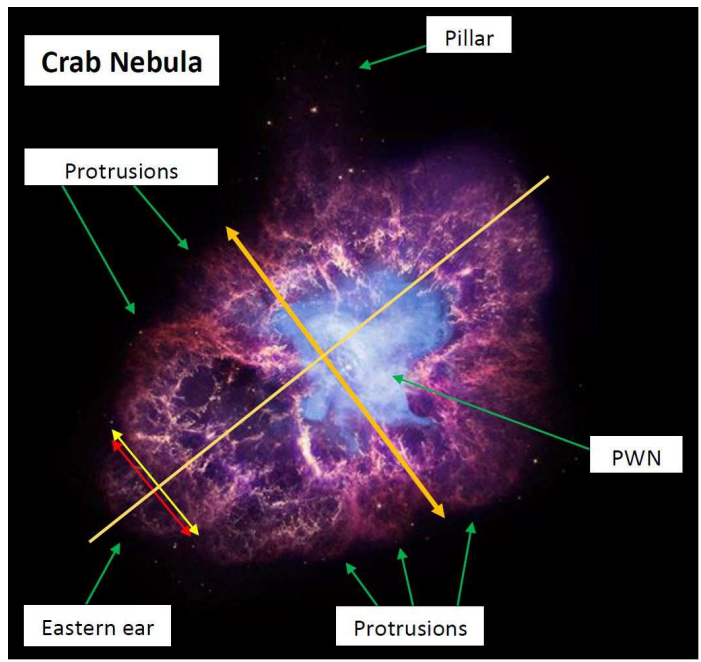}
\caption{The Crab Nebula. The composite image is assembled from X-ray (Blue), Optical (Red-Yellow) and Infrared (Purple). The image was taken from Chandra’s gallery. The X-ray is based on \citep{Sewardetal2006}, the Optical is based on \cite{Hester2008} and the Infrared is based on NASA/JPL-Caltech/Univ.
}
 \label{figure:CrabNebula}
\end{figure}

{\bf The Crab Nebula.} Fig.~\ref{figure:CrabNebula} presents one of the best studied SNR. Several protrusions are clearly resolved, including an eminent pillar. We can clearly identify only one ear, as marked on the figure, and revealed in the infra-red band. However, we note that it is barely noticed in the radio image \citep{Bietenholzetal2001}. The ear is qualitatively different from the other protrusions. It is along the long axis of the SNR, there is a sharp drop in surface brightness from the SNR shell to the ear, and the geometry of the ear is more spherical and not a continuation of the edge of the main SNR shell.  The diameter of the ear and the diameter of its base almost overlap because the   ear geometry is almost a half-sphere.

At the center of the Crab Nebula there is a pulsar with its PWN (e.g., \citealt{Hesteretal2002, Temimetal2006, Sewardetal2006}).
As mentioned by \cite{KomissarovLyubarsky2003} and \cite{Lyubarsky2012}, most of the energy of the pulsar is expelled in the equatorial plane, and not along the polar directions. A PWN might form two opposite ears by the pinching effect of the toroidal magnetic field (e.g., \citealt{BegelmanLi1992}), but as we discussed in section \ref{sec:jets}, we instead discuss the formation of ears by jets.

The age of a CCSN is an important parameter, as on average older CCSN are more likely to be influenced by the interstellar medium. 
As seen from Table \ref{table:age}, three of the CCSN are relatively young with an age of $<1000 \yr$. These SNRs are unlikely to have been influenced much by the ISM, and they support our assumption that the ears are formed by the SN progenitor before, during, or after the collision. The morphologies of evolved remnants (S147, Vela, W44 and even Puppis A) are more likely to have been be influenced by the ISM, but as we discussed earlier, we do not expect the ISM to form two opposite ears. We therefore attribute the presence of ears also in the older CCSNRe to jets, and the survival of the ears we attribute to the large energy that was deposited in inflating the ears. 
\begin{table}[H]
\begin{center}
 \begin{tabular}{||l c l||}
 \hline
 SNR & Approximated Age [years] & Source \\ [0.5ex]
 \hline
 Cassiopeia A & 350 & \cite{Ghiotto2015} \\
 3C58 &  835 & \cite{Kothes2016} \\
 Puppis A & 1990 & \cite{Aschenbach2015} \\
 S147 & 30,000 & \cite{Gvaramadze2006} \\ 
 Vela & 11,400 & \cite{Reichleyetal1970}, \cite{SushchHnatyk2014}  \\
 G309.2-00.6 & 4000 & \cite{Gaensleretal1998} \\
 W44 & 20,000 & \cite{Coxetal1999} \\
 Crab Nebula & 962 & \cite{PolcaroMartocchia2006} \\ [1ex]
 \hline
\end{tabular}
\centering
\caption{ The ages of the CCSNRs studied here. Ears are present in CCSNe with a range of ages. 
}
\label{table:age}
\end{center}
\end{table}

 \cite{Schencketal2014} detect a faint protrusion to the northeast of the CCSNR~B0049–73.6, and term it an ear. This protrusion does not comply with our definition of an ear. We therefore consider the presence of ears in this SNR as questionable.
There are three other CCSNRs that have general morphology that we consider as compatible with the presence of ears, but where we see no ears at present. These are SNR 0540-69.3  (that has a central pulsar), RCW~103 (that has a magnetar at its center),and W49B.
\cite{BearSoker2017} for example, compare the structure of SNR W49B with that of several planetary nebulae, some of them have ears. The CCSNR nature of W49B can be inferred from its mean metal abundances, that are consistent with the predicted yields in models of bipolar/jet-driven core-collapse SNe \citep{Lopezetal2013}. 

Together with the 8 CCSNe we presented in sections \ref{sec:jets} and \ref{sec:ears} (not including W50), we conclude that 8-12 out of the 28 CCSNRs with well resolved images have ears.
 This, together with projection effects that can hide ears in some CCSNRs, is the source of our estimate that at least third of CCSNRs have (or had) ears.

\section{ESTIMATING THE ENERGY TO INFLATE EARS}
\label{sec:energy}

We calculate the approximate fraction of the energy that was required to inflate each ear under the following assumptions.
(1) Before the jets formed the ears, the SNR shell was spherical. (2) As well, the mass density per unit area on the shell was constant. (3) The mass in the jet is small relative to the mass of the SNR that the jet interacted with and is now part of the ear. This holds if the velocity of the jets was much higher than that of the shell, as we assume here. (4) The ear was formed during, or shortly after, the explosion. (5) The ear has a more or less spherical shape. This is definitely not the case, but it is adequate to our goals when the other uncertainties are considered. (6) The velocity of the SNR shell and the ear did not change much since the formation of the ear.

Under assumption 2 the ratio between the mass in the ear, $M_{\rm ear}$, and the total mass in the SNR, i.e., including the ear mass, $M_{\rm SNR}$ , is
\begin{equation}
\mu \equiv
\frac {M_{\rm ear}}{M_{\rm SNR}}
= \frac{1- \cos \theta}{2}
=\frac{1}{2} \left[ 1 - \sqrt{ 1- \left( \frac{R_{\rm base}}{R_{\rm SNR}} \right)^2 } \right],
\label{eq:mu}
\end{equation}
where the radius of the SNR, $R_{\rm SNR}$, and the radius of the base of the ear, $R_{\rm base}$, are half the lengths of the double-headed thick orange arrows and yellow arrows in the images presented in sections \ref{sec:jets} and \ref{sec:ears}, respectively, and $\theta$ is the half opening angle of the ear as seen from the center of the SNR.

The average expansion velocity of the SNR and, under assumption 4, the average velocity of the ear are given by
\begin{equation}
v_{\rm SNR} =\frac{R_{\rm SNR}}{t},
\label{eq:VSNR}
\end{equation}
and
\begin{equation}
v_{\rm ear} =\frac{R_{\rm SNR}+R_{\rm ear}}{t},
\label{eq:Vear}
\end{equation}
where $t$ is the age of the SNR, and the radius of the ear, ${R_{\rm ear}}$, is half the length of the double-headed red arrows in the figures presented in section \ref{sec:jets} and \ref{sec:ears}.

Under our assumptions the kinetic energy of the gas that is now in the ear before it was acted upon by the jet is given by
\begin{equation}
E_{\rm ear,0} \simeq \frac{M_{\rm ear}v_{\rm SNR}^2}{2},
\label{eq:Eear0}
\end{equation}
and the present kinetic energy of this gas is
\begin{equation}
E_{\rm ear} \simeq \frac{M_{\rm ear}v_{\rm ear}^2}{2}.
\label{eq:Eear}
\end{equation}

We assume that the SNR moves supersonically, and we neglect the thermal energy of the gas. We now make the assumption that the entire energy of the jets was transferred to the extra energy of the ears. This is a strong  assumption, as during the interaction, dissipation heats the gas and some of the energy is radiated away. This assumption, therefore, underestimates the energy of the jet. Subtracting the kinetic energy of the gas in the ear before the interaction with the jet, $E_{\rm ear,0}$, from the kinetic energy of the ear, $E_{\rm ear}$, gives an estimate of the energy that was required to inflate the ear by the jet
\begin{equation}
E_{\rm jet} \approx \Delta E_{\rm ear}= E_{\rm ear}-E_{\rm ear,0}.
\label{eq:deltaE1}
\end{equation}
The desired quantity is the ratio, $\epsilon _{\rm ear}$, of this energy to that of the entire SNR shell,
\begin{equation}
E_{\rm SNR} \simeq  \frac{M_{\rm SNR}v_{\rm SNR}^2}{2}.
\label{eq:ESNR}
\end{equation}
We find
\begin{equation}
\epsilon_{\rm ear} \equiv  \frac{\Delta E_{\rm ear}}{E_{\rm SNR}}
  \simeq  \mu \left[\left(1+\Gamma\right)^2 - 1\right],
\label{eq:epsilon1}
\end{equation}
where we defined
\begin{equation}
\Gamma \equiv
\frac {R_{\rm ear}}{R_{\rm SNR}}.
\label{eq:Gamma}
\end{equation}

In Table \ref{table:calculations} we present the results of our calculations, for each ear separately and for the two ears combined ($\epsilon_{\rm ears}$) for each SNR.
\begin{table}[H]
\begin{center}
 \begin{tabular}{||c c c c||}
 \hline
 SNR & $\epsilon_{\rm west}$ & $\epsilon_{\rm east}$ & $\epsilon_{\rm ears}$ \\ [0.5ex]
 \hline
 Cassiopeia A & 0.038 & 0.064 & 0.10 \\
 3C58 & 0.037 & 0.032 & 0.07 \\
 Puppis A & 0.009 & 0.010 & 0.02 \\
 S147 & 0.039 & 0.072 & 0.11 \\
 Vela & 0.005 & 0.004 & 0.01 \\
 G309.2-00.6 & 0.039 & 0.03 & 0.07 \\
 W44 & 0.034 & 0.029 &  0.06 \\
 Crab Nebula & - & 0.034 & 0.03 \\ [1ex]
 \hline
\end{tabular}
\centering
\caption{The ratio of the extra energy in the ears to that in the SNR shell. The last column is the combined values of the two ears. }
\label{table:calculations}
\end{center}
\end{table}

We estimate the energy of the jets that inflated the eastern ear of Cassiopeia A to be 6.4 per cents of the explosion energy. 
\cite{Lamingetal2006} estimate this jet energy to be about $10^{50} \erg$. For the explosion energy of the entire supernova they take $2-4 \times 10^{51} \erg$. So their estimate is that the jet to SN energy ratio is about $2.5-5$ per cents. 
\cite{Orlandoetal2016} perform numerical simualtions to match the morphology of Cassiopeia A. They take the explosion energy to be $2.3 \times 10^{51} \erg$, and estiamte the energy of the eastern and western ears to be $1.8 \%$ and $0.4 \%$ of the exploison energy, respectively.
\cite{FesenMilisavljevic2016} estimate the kinetic energy of the eastern ear to be $\approx 10^{50} \erg$, more than twice the enrgy estaimted by \cite{Orlandoetal2016}, i.e., about $4 \%$ of the exploison energy.    
Considering the many uncertainties, we consider our value to be satisfactorily close to the values listed above.

In the interaction of the jets with the shell, some energy might be radiated away or be spread in a larger volume than that of the ear (e.g., to the sides of the base of the ear). Our calculation does not take these effects into account. For that, the energy of the inflating jets is somewhat larger than the extra energy in the ear $E_{\rm jet} \ga \Delta E_{\rm ear}$.
As we mentioned in section \ref{sec:ears}, \cite{Micelietal2013} estimated the kinetic energy of the eastern ear in Vela SNR (i.e. the clump) to be about $5 \%$ of the kinetic energy of the SNR shell. According to our calculation the extra kinetic energy of the eastern ear is about $2 \%$ of the total energy of the SNR. This demonstrates that the energy in the jets might be about a factor of $\approx 2$ larger than the estimated energy in the ears (derived under different assumptions).
The time of interaction between the jets and the SNR shell determines not only the amount of energy that does not end in the ear, but also the the exact shape of the ear. Both quantities will have to be determined by a set of numerical simulations.

At this point we only make a crude estimate. Based on the value of $\epsilon_{\rm ears} \approx 0.01-0.1$ we state that the typical energy in the two opposite jets is
\begin{equation}
\epsilon_{\rm jets} \equiv  \frac{E_{\rm jets}}{E_{\rm SNR}}\approx 0.05-0.15
\label{eq:epsilonjets}
\end{equation}

\section{DISCUSSION AND SUMMARY}
\label{sec:Summary}

We estimated the extra kinetic energy that was required to inflate ears in our sample of eight CCSNRs, under the assumption that they were inflated by jets. Our definition for ears is given at the beginning of section \ref{sec:jets}.
We calculated the extra energy for eight CCSNRs whose images are presented in sections \ref{sec:jets} and \ref{sec:ears}. For seven CCSNRs we identified two clear opposite ears, whereas for the Crab Nebula we could preform the calculation for only one well identified ear.

The ears can in principle be formed before the explosion, if the SN expands to a CSM with ears, during the explosion, or after the explosion. As well, older SNRs are more likely to be influenced by the ISM, but we do not believe the ISM can form two opposite ears. We here made the assumption that the ears have been shaped during the explosion, or shortly after it. This holds true in at least some CCSNRs. For example, it seems that the Crab Nebula does not interact with a dense CSM or ISM (e.g., \citealt{YangChevalier2015}). In the Cassiopeia A SNR there is an observational indication for a jet that inflated the ear (\citealt{Hwangetal2004, Lamingetal2006}; see discussion in section \ref{sec:jets}). However, without detailed modeling of each remnant, we cannot be certain how the ears were formed.

We crudely estimated the energy that is required to inflate the ears in 43 $\%$ of CCSNRs from our original sample, and presented the results in Table \ref{table:calculations}. As not all the energy of the jets ends in the ears, the original energy of the jets might be somewhat larger. Over all, we crudely estimate that the original energy in the two opposite jets that inflate ears in CCSNe is a fraction of $\epsilon_{\rm jets} \approx 5-15 \%$ of the kinetic energy of the SNR shell (eq. \ref{eq:epsilonjets}).

Based on the eight CCSNRs where we identified ears (not including W50), and four questionable  CCSNRs, we estimated the fraction of CCSNRs which posses opposite ear-like features (section \ref{sec:ears}) to be $\approx 29-43 \% $ of CCSNRs. If the symmetry axis, i.e., the line joining the two counter ears, is close to our line of sight, the ears will be projected on the main SNR shell and we might not see them. Therefore, the true fraction of CCSNRs with ears is likely to be higher, at a value of at least third of all CCSNRs, and more likely $\approx 40 \%$ of of all CCSNRs.

We propose that the jet feedback mechanism (JFM; e.g., \citealt{PapishSoker2011, GilkisSoker2016, Soker2016}) is compatible with the large fraction of CCSNRs with jet-inflated ears, and with a typical energy relative to the explosion energy of $\epsilon_{\rm jets} \approx 5-15 \%$.  
The energy of the jets of $\epsilon_{\rm jets} \approx 5-15 \%$ is explained by the feedback explosion process. 

The process we proposed here, of jets launched during the explosion or shortly after it, is similar in some respects to jets that shape ears in some planetary nebulae. In planetary nebulae the jets might be formed before or after the main nebular shell (e.g., \citealt{Tocknelletal2014}). In those planetary nebulae with ears similar to those of SNRs, it is possible that the age of the jets and nebular shells is similar, and the jets have played a role in ejecting the nebula, possibly in a feedback process (e.g., \citealt{Shiberetal2016}).
We hence speculate that the formation processes of ears in planetary nebulae might have some similarity to the formation of ears in CCSNRs, and in any case, both processes involve energetic jets. Based on our study of ear features in CCSNRs, we suggest that the JFM merits serious consideration as a mechanism for producing CCSNRs. We anticipate that further theoretical and observational studies will place this promising mechanism on increasingly sure footing.
 
 We thank an anonymous referee for very useful comments. 
A.G. was supported by The Rothschild Scholars Program- Technion Program for Excellence.  We Acknowledge support by the E. and J. Bishop Research Fund at the Technion.

\end{document}